# Development of Multi-Filament Textured-Powder Bi-2212/Ag wire with enhanced LAR

P.M. McIntyre *Member IEEE,* T. Elliott, G.D. May, J.S. Rogers, and A. Sattarov

*Abstract* The performance and cost of Bi-2212/Ag wire is limited by the large fraction of silver matrix (~4:1) that is required in the oxide-powder-in-tube fabrication process. An alternative fabrication process is being developed in which fine-powder Bi-2212 is uni-axially compressed to form bars with a thin Ag foil sheath. The fine powder naturally textures (aligns the a-b planes perpendicular to the direction of compaction) with texture >80% using 200 MPa compression.

A billet is formed by stacking trapezoidal-cross-section bars in a symmetric 8-12-16 pattern around a Ag rod and enclosing in a Ag-wall extrusion can. The billet is extruded and drawn to fine wire.

Results are presented on present status of the development and testing.

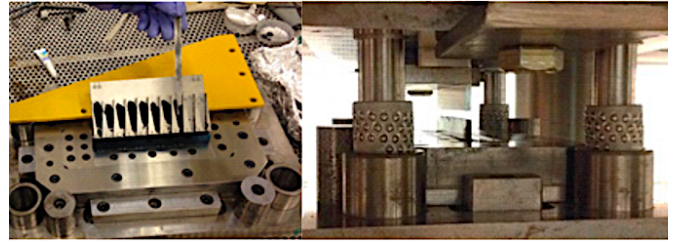

Fig. 1. Fabrication of square-cross-section textured-powder Bi-2212 core: a) load fine-powder Bi-2212 into rectangular die; b) compress to 200 MPa.

## I. Introduction

**B**i-2212 is the only one of the Type II high-temperature superconducting ceramics that has been successfully fabricated as a multi-filament round wire [1]. The perovskite superconductors have a layered crystal structure in which bonding at the $CuO_2$ layer is weakest. The $CuO_2$ planes convey two important properties. First, below the critical temperature superconducting current flows preferentially in the (a-b) $CuO_2$ planes – supercurrent transport is strongly anisotropic. Second, crystalline Bi-2212 is micaceous. If a single crystal is ground to a powder, the individual grains of the powder have very flat aspect ratio – each grain is much larger in the a-b plane than in the c-direction. Also like mica, the particles of Bi-2212 powder slide freely when they are oriented with a-b planes parallel.

Bi-2212 is conventionally fabricated into multi-sub-element wire using an oxide-powder-in-tube (OPIT) method [2] in which a fine powder of Bi-2212 is poured into the aperture of a silver tube, the tube ends are plugged, and the tubes are assembled into a hex-close-packed (HCP) registration. The assembly is drawn, re-stacked and re-drawn to form a multi-sub-element composite wire. The drawing properties of the Bi-2212 cores and the silver matrix are significantly different. The tap density of the powder must be controlled in filling the tubes, otherwise the powder can aggregate during the area-reducing draw and 'lock up' to produce sub-element breaks and sausaging during successive stages of drawing. The billet geometry for OPIT-method wire must be limited to a local area ratio (LAR) Ag:Bi-2212 ~4:1 so that the silver matrix controls the area-reducing draw to sustain the registration of the Bi-2212 powder cores within the silver matrix as the wire is drawn. Because of this limit on LAR, the cost of the silver content of Bi-2212/Ag wire is comparable to that of the Bi-2212 powder, and the overall wire cost is problematically high for many applications in superconducting devices.

After a multi-sub-element wire is completed, and cabled, and then wound into a winding for a practical application, the winding must be heat-treated in its final shape to the melt temperature of Bi-2212 (~880 C) in an oxygen-rich atmosphere. The Bi-2212 powder within each sub-element is melted within its Ag channel, and then the temperature is lowered gradually to re-crystallize and anneal the Bi-2212 in dendritic crystallites. The temperature is slowly lowered to anneal the grain structure within each sub-element, and indeed to grow the Bi-2212 grains to interconnect through the silver matrix among neighboring sub-elements within the wire.

In a further important improvement to wire processing [3], the heat treatment is performed in a high-pressure retort, containing ~50 bar Ar and 1 bar $O_2$. The high pressure collapses the void spaces in the solid cores and in the melt liquid, so that recrystallization and annealing are done preserving full density. Over-pressure (OP) processing more than doubled the attainable superconducting current density in the wire.

(Style: TAS First page footnote) Manuscript receipt and acceptance dates will be inserted here. This research was supported by SBIR grant DE-SC0020712 from the US Dept. of Energy.

P.M. McIntyre is with the Accelerator Technology Lab, Texas A&M University, College Station, TX 77845 USA and also with Accelerator Technology Corp. (e-mail: p-mcintyre@tamu.edu).

G.D. May and J.S. Rogers are with Accelerator Technology Lab, Texas A&M University, College Station, TX 77845 (email: jsr12e@tamu.edu).

A. Sattarov is with Commonwealth Fusion Systems, Boston, MA USA (e-mail: akhdiyorsattarov@gmail.com).

T. Elliott is with Accelerator Technology Corp., College Station, TX 77845 (email: elliott@tamu.edu ).

Color versions of one or more of the figures in this paper are available online at http://ieeexplore.ieee.org.

Digital Object Identifier will be inserted here upon acceptance.

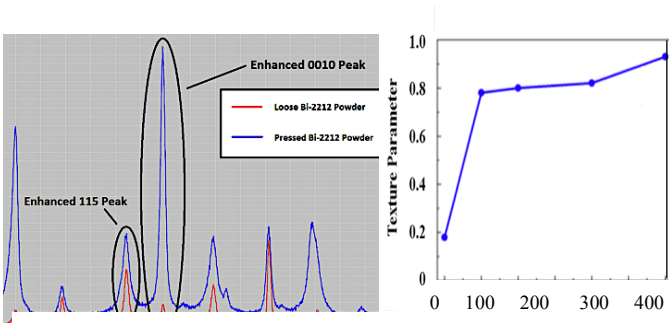

Fig. 2. a) XRD spectrum of exposed face in Bi-2212 bar: before and after compression; b) texture parameter as a function of compression.

In a previous development[4] it was found that a quantity of Bi-2212 fine powder can be textured by uniaxial compression to produce beneficial properties for sub-element cores. The uniaxial compression naturally orients the a-b planes of the powder grains so that they are normal to the compression axis – the compression produces a textured-powder bar. The procedure is illustrated in Fig. 1: a quantity of Bi-2212 fine powder is placed in a rectangular channel in a compression die and a conformal male die is placed on top of the powder so that the powder is confined within the rectangular channel. The male die is pressed into the channel using a hydraulic press, so that the powder is uniformly compressed in the direction normal to the axis of the channel. The compression also produces a cold-sintering of the powder so that the bar is weakly bound in its shape once removed from the rectangular channel.

Experiments [4] were performed in which textured-powder (TP) bars are fabricated, the bar is split along the texturing plane (normal to the direction of compression and to the bar axis) to expose a layer of the bar's interior, and the split sample is mounted for analysis of the morphology and texturing of the powder particles. X-ray diffraction (XRD) is performed using the Bragg-Brentano method [5] to measure the texture parameter $\tau$ is extracted from the signal strengths of those two peaks, weighted by their oscillator strengths. Fig. 2a shows the XRD spectrum of one sample, in which specific diffraction directions are identified that are produced by diffraction from crystallites with an ordered a-b plane (0010) and with completely disordered a-b planes (115):

$$\tau = \frac{I(0010) - 0.25 I(115)}{I(0010) + 0.75 I(115)} \quad (1)$$

Fig. 2b shows the dependence of $\tau$ on compression: a texture of >70% could be achieved when Bi-2212 powder of particle size ~1 µm is compressed into 4 mm square-cross-section TP bars with a compressive stress of ~200 MPa at room temperature. The bars were split at multiple locations through the thickness of the bar to assess the degree to which texturing was homogeneous. Very little variation of $\tau$ was observed, indicating that texturing is homogenous throughout a 4 mm square bar.

The TP bars were assembled into a copper-clad silver billet (Figure 2c) and extruded and then drawn to fine wire. Studies of the drawn wire showed that the textured powder within the bar has sliding friction during an area-reducing draw that is much less than that of an un-textured filling of Bi-2212 powder within a round tube, so that it draws uniformly without sausaging or breaks. This discovery motivated the present work, in which an optimized billet geometry is designed with the objective of reducing the local area ratio (LAR) = Ag:Bi-2212 from the value of 4:1 typical of OPIT wire to a target value of ~1:1.

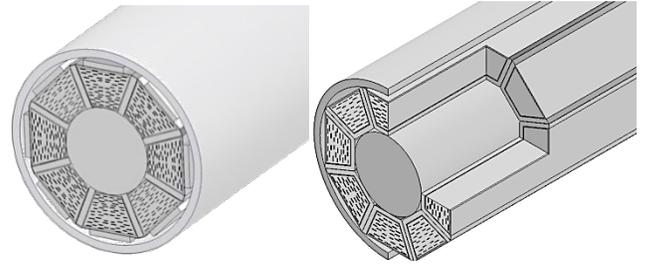

Fig. 3. Isometric cutaway view of single-layer billet design for textured-powder core (8 sub-elements; 14.5 mm OD; LAR = 2.1).

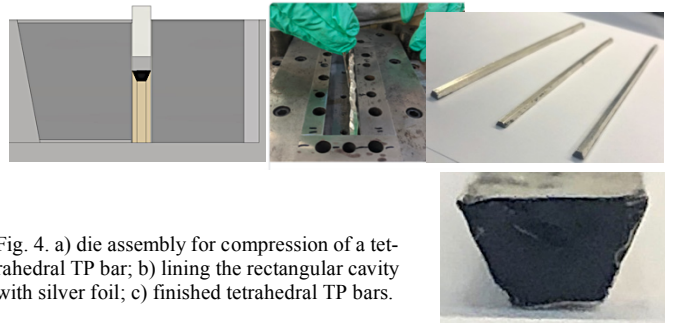

Fig. 4. a) die assembly for compression of a tetrahedral TP bar; b) lining the rectangular cavity with silver foil; c) finished tetrahedral TP bars.

## II. BILLET GEOMETRY

The billet geometry for assembling Bi-2212 TP bars should have an axial symmetry of bars in the space between a round Ag core rod and a cylindrical Ag alloy can. The geometry is further constrained to orient the textured a-b planes of the bars so that they are azimuthal around the axis of the billet. As the billet is drawn down to fine wire, it is crucial that the interior registration of the pattern and shape of the cores should be preserved.

Those considerations led to the choice of a tetrahedral cross-section for the TP bars. Fig. 3 shows a single-layer billet in which 8 tetrahedral bars are configured around a core rod of pure silver. The angle between the short inner face of each tetrahedral bar and each of its side faces is 45°, chosen so that the faces of neighboring bars are parallel.

The method for forming the tetrahedral bars is shown in Fig. 4. The cavity in the female die is lined with a Ag foil strip (50 µm thick). The Bi-2212 loose powder is filled in 8 equal portions along the slot and equalized in depth. The top flaps of silver foil are folded over the powder filling, and the male die is installed. The silver foil is compressed into all surfaces of the bar, and also provides a slip-plane between the powder and the die. After compression the TP bars releases readily from the die and is ~hermetically sealed.

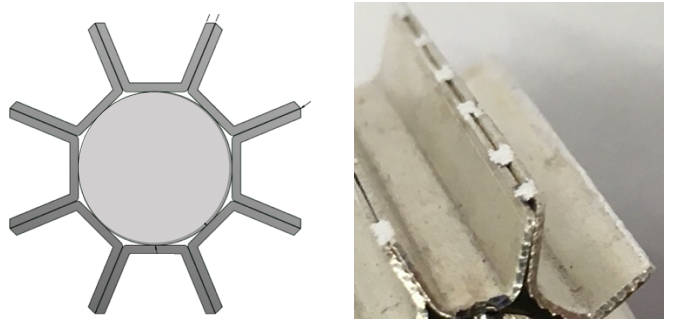

Fig. 5. a) Array of silver alloy U-channels, assembled onto silver core rod; b) laser spot-weld to join the outer edges of the side strips of neighboring U-channels to stabilize the array.

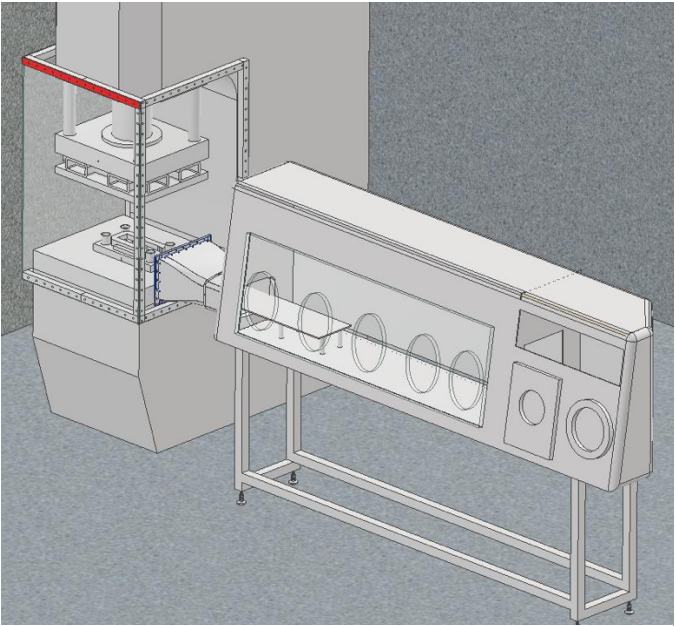

Fig. 6. Glove box and hydraulic press used to fabricate TP bars and assemble billets.

Two methods for assembling the TTP bars in the billet are being evaluated, as shown in Fig. 3. In the first method, strips of Ag alloy foil are interspersed with the bars in the billet geometry. In the second method, a U-channel is coined from Ag alloy foil in the geometry to conform to 3 sides of the TP bar, as shown in . The U-channels are assembled around the core bar and spot-welded at locations along their common edges using laser welding (Fig. 5).

The entire operation of filling powder into the die assembly, compressing the TP bars, and assembling the bars into a Ag billet is performed in a clean environment with inert gas purge flow. A glove box has been integrated with a hydraulic press for the purpose, as shown in Fig. 6. A program of development is under way to determine the optimum billet configuration.

### III. PROCESS DEVELOPMENT PLANS

Single-layer billets are now being fabricated, in which the above assemblies will be inserted into a Ag alloy can and sealed with Ag end plugs. Two methods for area reduction will be investigated: extrusion using an indirect extrusion press at Ames Laboratory followed by drawing; and pointing and drawing without the extrusion step. Samples will be sliced after each area reduction and analyzed for the fidelity of registration of the sub-elements and any evidence of sausaging or sub-element breaks. Wire will be drawn to 1 mm diameter, and samples will be OP processed and tested for $I_c$ at NHMFL.

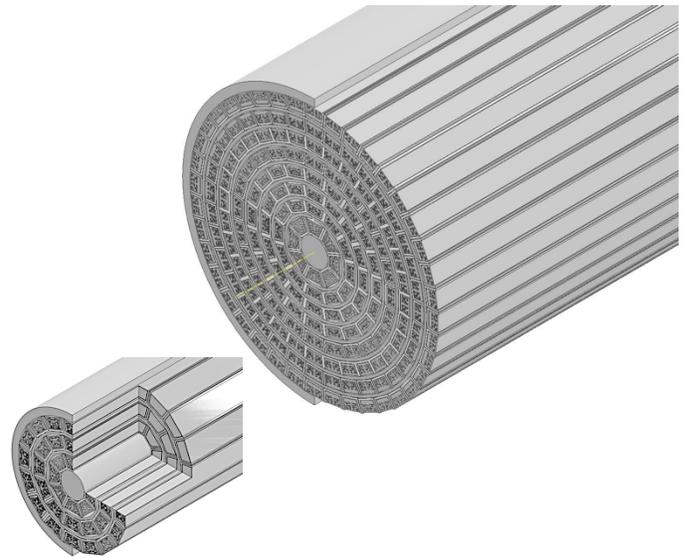

Fig. 7. 3-layer billet using the same billet geometry (27 mm diameter, 36 sub-elements; LAR = 1.25); b) 8-layer billet (61 mm diameter, 176 sub-elements; LAR = 0.95).

### IV. SuperCIC, OP HEAT TREATMENT

Once TP wire fabrication is optimized using single-layer R&D billets, the process will be scaled to 3-layer billets and to 8-layer billets following the strategy summarized in Fig. 7. Providing the structural design supports extrusion and drawing to fine wire without sub-element breakage or sausaging, the TP strategy should yield a LAR ~1:1.

Several new routes have been developed for synthesis of high-purity single-phase fine powders of Bi-2212 [6]. The new powders yield even greater $I_c$ performance. A continuing challenge for wire cost, however, is that the new powders are made using high-purity primary chemicals in order to eliminate possible contaminants that might promote secondary phase formation or otherwise compromise performance. It is anticipated that the TP bars will provide a stable precursor form for the fabrication of billets and area-reduction to fine wire. As one pathway in process development, it is planned to undertake a parallel development in which lower purity (and significantly lower-cost) specifications for the primary chemicals will be used in trial batches of powder. The objective is to develop relaxed but sufficient standards for ingredient purity that can significantly reduce powder cost.

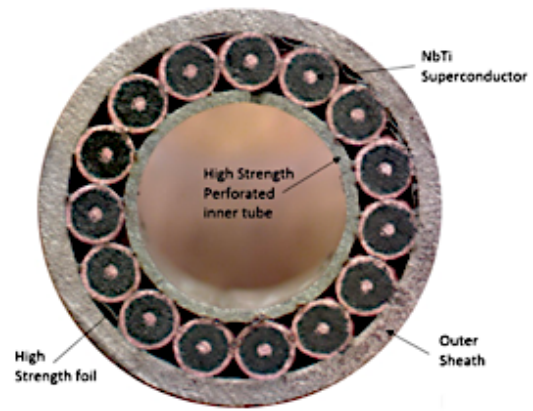

Fig. 8. SuperCIC conductor, fabricated using NbTi wire. The same conductor could be fabricated using JR-process Bi-2212/Ag wire.

The utilization of TP Bi-2212/Ag wire for applications to high-field magnets, generators, and other devices will require a cable and coil technology that can support both the *in situ* OP heat treatment of final-geometry windings and stress management at the cable and sub-winding level that can support fabrication and assembly of hybrid windings. ATC and TAMU have developed a SuperCIC cable [7], shown in Fig. 8, in which superconducting wires are cabled with twist pitch onto a perforated center tube, the cable is pulled as a loose fit into a sheath tube, and the sheath is drawn down onto the cable to compress the wires against the center tube and immobilize them. SuperCIC provides a robust basis for coil technology, since the sheath provides stress management at the cable level, formation of compact end windings, and fabrication and separate heat treatment of sub-winding assemblies.

SuperCIC provides a particularly interesting basis for windings utilizing TP Bi-2212 wire. The outer sheath tube could be made from a recently developed high-aluminum-content Ni-based superalloy (Haynes 233) that forms a self-passivating diffusion barrier and also sustains high strength at the 880 C heat treatment required for OP processing. The sheath tube could itself provide safe containment of 50 bar, so OP heat treatment could be performed in a standard furnace. This benefit would be of importance for fabrication of practical Bi-2212 windings and sub-windings.

## CONCLUSIONS

Textured-powder bars of Bi-2212 have been successfully fabricated by uniaxial compression. 70% texture can be attained by 200 MPa compression, and when the TP bars are compounded into a billet and drawn, shear friction is reduced so the billet can be drawn to fine wire with no sausaging or sub-element breaks. Multi-sub-element billets are being fabricated using tetrahedral-cross-section TP bars, with the objective of developing a next-generation fine-filament Bi-2212 wire with LAR ~1:1.

## ACKNOWLEDGMENTS

Matthew Besser and his team at Ames Laboratory have prepared and rolled Ag alloy tubes for the extrusion billet, and will extrude billets using their indirect extrusion press. Jianyi Jiang and his colleagues at NHMFL plan to perform OP heat treatment, critical current testing, and microstructure analysis on samples of TP Bi-2212 wire. The authors thank Mike Tomsic (HyperTech Research) for many helpful discussions and suggestions about process development.